\documentclass[pra,twocolumn,superscriptaddress,showpacs]{revtex4-1}

\usepackage{graphicx}
\usepackage{dcolumn}
\usepackage{bm}

\newcommand{\beq}{\begin{equation}}
\newcommand{\eeq}{\end{equation}}
\newcommand{\beqa}{\begin{eqnarray}}
\newcommand{\eeqa}{\end{eqnarray}}
\newcommand{\Tr}{\text{Tr}}

\newcommand{\half}{\frac{1}{2}}

\newcommand{\ket}[1]{\left | #1 \right \rangle}

\newcommand{\kket}[1]{| #1 \hspace{-.9mm}\left. \right>}
\newcommand{\bbra}[1]{  \hspace{-.6mm}\left<\hspace{-.4mm} \right. #1 |  }

\newcommand{\ii}{\mathrm{i}}
\newcommand{\ee}{\mathrm{e}}

\newcommand{\bz}{k_{\text{B}}}

\begin{document}


\title{Thermodynamic Work Gain from Entanglement}


\author{Ken Funo}
\affiliation{Department of Physics, The University of Tokyo, 7-3-1 Hongo, Bunkyo-ku, Tokyo, 113-0033, Japan}
\author{Yu Watanabe} 
\affiliation{Yukawa Institute for Theoretical Physics, Kyoto University, Kitashirakawa Oiwake-Cho, Kyoto, 606-8502, Japan}
\author{Masahito Ueda}
\affiliation{Department of Physics, The University of Tokyo, 7-3-1 Hongo, Bunkyo-ku, Tokyo, 113-0033, Japan}

\date{\today}

\begin{abstract}
We show that entanglement can be utilized to extract the thermodynamic work beyond classical correlation via feedback control based on measurement on part of a composite system. The net work gain due to entanglement is determined by the change in the mutual information content between the subsystems that is accessible to the memory.
\end{abstract}


\maketitle

\section{\label{sec:Intro}Introduction}

Feedback control of thermal fluctuations has been discussed by a number of researchers~\cite{JMaxwell,Szilard,Brillouin,Landauer,Bennett,Maxwell,Rev,Sagawa1,Sagawa2, Sagawa3,Sagawa5,Jacobs,Toyabe,Berut} since the seminal work by Maxwell~\cite{JMaxwell} who pointed out that if an observer can access microscopic degrees of freedom, the second law of thermodynamics may break down. Szilard devised a quintessential model of Maxwell's demon, in which $\bz T\ln 2$ of work can be extracted from a thermodynamic cycle~\cite{Szilard}. It has been shown~\cite{Sagawa1,Sagawa2} that extractable work $W^{S}_{\text{ext}}$ from a system is determined by the information gain (or the quantum-classical mutual information) $I$~\cite{Ozawa,Groenewold,Sagawa1} which measures the acquired knowledge of the system via measurement, and that the same quantity sets the lower bound on the total cost $W^{M}_{\text{cost}}$ of measurement and erasure of information:
\beqa
W_{\text{ext}}^{S} &\leq& -\Delta F^{S}+k_{\mathrm{B}}TI  , \\
W_{\text{cost}}^{M} &\geq & k_{\mathrm{B}}TI,
\eeqa
where $\Delta F^{S}$ is the free-energy difference of the system. It has also been shown that a positive entropy production in measurement compensates for the work gain via feedback control~\cite{Sagawa3}. Quite recently, Maxwell's demon~\cite{Toyabe} and Landauer's principle~\cite{Berut} have been demonstrated experimentally.

There has been considerable experimental interest in quantum feedback control concerning, for example, cooling~\cite{Bannerman,Baugh,Bushev} and spin squeezing~\cite{Fernholz}.
A number of results have been obtained that exploit quantum entanglement and cannot be achieved classically~\cite{Hotta,Lutz,Hormoz,Zurek,Brodutch,quantummemory,Partovi,Jennings,Horodecki,Oppenheim,Tajima,Kim}. Entangled states can also be used to extract energy by controlling quantum fluctuations via feedback control~\cite{Hotta}. The difference in the extractable work between local and nonlocal Maxwell's demons is closely related to quantum discord~\cite{Zurek,Brodutch}. In Refs.~\cite{Horodecki,Oppenheim}, work gain from entangled states using LOCC protocols is discussed. However, the ability of genuine quantum entanglement to obtain thermodynamic work gain has yet to be fully explored. In this paper, we address the question of whether quantum entanglement can be utilized as a resource for the thermodynamic work gain beyond classical correlation. We show that work can be extracted by feedback control from quantum correlations between the systems as well as from their thermal fluctuations.

To examine the effect of measurement and feedback control, let us consider a qubit system. Let the initial state be the maximally mixed state, i.e., $\rho=\frac{1}{2}(\kket{\downarrow}\bbra{\downarrow}+\kket{\uparrow}\bbra{\uparrow})$. The von Neumann entropy of the initial state is given by $S(\rho)=\ln2$, where $S(\rho):=-\Tr[\rho\ln\rho]$. Now let us perform a projective measurement using the basis $\{\kket{\downarrow},\ \kket{\uparrow}\}$. Depending on the measurement outcome, we perform a unitary operation $U_{\downarrow}=1$ and $U_{\uparrow}=\kket{\uparrow}\bbra{\downarrow}+\kket{\downarrow}\bbra{\uparrow}$, which flips the spin if the measurement outcome is $\uparrow$. Then, the final state is given by a pure state irrespective of the measurement outcome: $\rho'=\kket{\downarrow}\bbra{\downarrow}$ and the entropy of the vanishes, i.e., $S(\rho')=0$. From the thermodynamic relation; $F=E-TS$, we observe that the average reduction in entropy via measurement and feedback leads to the free-energy gain, which, in turn, can be extracted from the system as work.

The above protocol can be generalized by considering a general quantum measurement on the system by introducing a memory and performing an indirect measurement, i.e. a unitary transformation on the composite system followed by a projective measurement on the memory~\cite{Nielsen}. Also, the feedback control is realized by performing a unitary transformation on the system depending on the measurement outcome as exemplified above.

This paper is organized as follows. In Sec.~\ref{sec:Setup}, we introduce a protocol to realize the measurement and feedback control on entangled states. In Sec.~\ref{sec:slaw}, we derive a generalized second law of thermodynamics under the protocol discussed in Sec.~\ref{sec:Setup}. We give an upper bound for the work that can be extracted from an entangled resource. An essential distinction from the classical case is that the classical mutual information is replaced by the quantum mutual information. Since the latter can, in general, be larger than the former due to entanglement, the upper bound of the extractable work can also be larger than the classical case. A similar result is obtained for the measurement cost. In Sec.~\ref{sec:discussion}, we summarize the main results of this paper and discuss future work.

\begin{figure}[b]
\includegraphics[width=.35\textwidth]{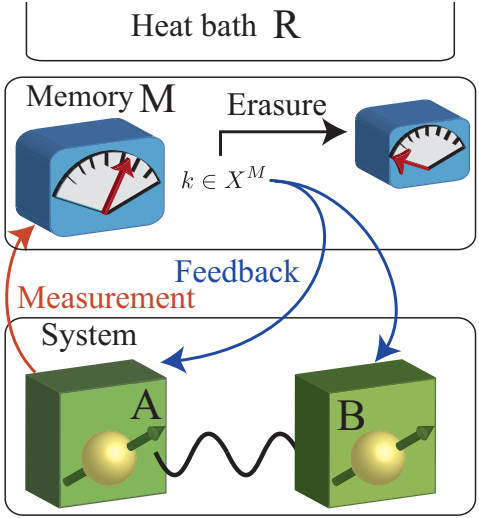}
\caption{\label{fig:epsart} (color online). Schematic illustration of our setup. We perform a measurement on subsystem $A$ of an initially entangled composite system $AB$, and obtain an outcome $k$. We then perform a feedback control on each subsystem via a local unitary operation that depends on $k$. The thermodynamic cycle is completed with the initialization or erasure of the memory.}
\end{figure}

\section{\label{sec:Setup}Setup}

We consider a composite system $AB$ and a memory $M$ that stores information about the measurement outcome. We consider the following protocol (see Fig.~\ref{fig:epsart}).

1. Let the initial Hamiltonian of the total system be $H_{\mathrm{i}}=H^{A}_{\mathrm{i}}\otimes H^{B}_{\mathrm{i}}\otimes H^{M}$. Here, we decompose the Hilbert space of the memory into mutually orthogonal subspaces $\bf{H}^{M}=\bigoplus_{k}\bf{H}^{M}_{k}$ according to the measurement outcome $k$~\cite{Sagawa2}. We also decompose the Hamiltonian of the memory as $H^{M}=\oplus_{k}H^{M}_{k}$, so that each support of $H^{M}_{k}$ belongs to $\bf{H}^{M}_{k}$. We assume that the initial state of the memory is the standard state belonging to $\bf{H}^{M}_{0}$. Let the initial state be $\rho_{\mathrm{i}}^{ABM}=\rho_{\mathrm{i}}^{AB}\otimes\rho^{M}_{0}$. The initial correlation between subsystems $A$ and $B$ is characterized by the quantum mutual information: 
\beq
I(\rho_{\mathrm{i}}^{A}:\rho_{\mathrm{i}}^{B}):=S(\rho_{\mathrm{i}}^{A})+S(\rho_{\mathrm{i}}^{B})-S(\rho_{\mathrm{i}}^{AB}). \label{quantummutual}
\eeq

2.  A general quantum measurement on the subsystem $A$ is implemented by performing a unitary transformation $U^{AM}$ on $AM$ followed by a projection measurement $P^{M}_{k}=\sum_{b}|\psi^{M}_{k}(b)\rangle\langle\psi^{M}_{k}(b)|$ on $M$, where $\{|\psi^{M}_{k}(b)\rangle\}$ is an orthogonal basis set in $\bf{H}_{k}^{M}$. The measurement outcome $k$ is registered by the memory during this process. The postmeasurement state depending on the outcome $k$ is given by 
\beqa
& &\rho^{ABM}(k)\nonumber \\
:=& &\frac{1}{p_{k}}P^{M}_{k}U^{AM}(\rho_{\mathrm{i}}^{AB}\otimes\rho^{M}_{0})U^{\dagger AM} P_{k}^{M} \nonumber \\
=& &\frac{1}{p_{k}}\sum_{a,b,c}M_{k,a,b}^{A}\rho^{AB}_{\mathrm{i}}M^{\dagger A}_{k,a,c}\otimes|\psi^{M}_{k}(b)\rangle\langle\psi^{M}_{k}(c)|,
\eeqa
where
\beq
p_{k}=\Tr[P^{M}_{k}U^{AM}(\rho_{\mathrm{i}}^{AB}\otimes\rho^{M}_{0})U^{\dagger AM} P_{k}^{M}]
\eeq
is the probability of obtaining $k$. Here 
\beq
M^{A}_{k,a,b}:=\sqrt{p^{M}_{0}(a)}\langle\psi^{M}_{k}(b)|U^{AM}|\psi^{M}_{0}(a)\rangle
\eeq
is a measurement operator acting on $A$, where
\beq
\rho^{M}_{0}:=\sum_{a}p^{M}_{0}(a)|\psi^{M}_{0}(a)\rangle\langle \psi^{M}_{0}(a)|,
\eeq
and it satisfies the relation
\beq
\sum_{k,a,b}M^{A\dagger}_{k,a,b}M^{A}_{k,a,b}=1.
\eeq

We note that the reduced density matrix of $AB$ is given by
\beq
\rho^{AB}(k)=\frac{1}{p_{k}}\sum_{a,b}M^{A}_{k,a,b}\rho^{AB}_{\mathrm{i}}M^{\dagger A}_{k,a,b}. \label{postmesstateab}
\eeq

3. We detach the memory from $A$, and perform a feedback control on each subsystem by means of local unitary operators $U^{A}_{k}$ and $U^{B}_{k}$ that depend on $k$. The unitary operators are given by $U^{A(B)}_{k}:=\text{T}\exp[-\ii\int^{t_{\mathrm{f}}}_{0}H_{k}^{A(B)}(t)\mathrm{d}t]$, where $\text{T}$ is the time-ordering operator and $H^{A(B)}_{k}(t)$ is the Hamiltonian of $A(B)$ at time $t$ depending on the measurement outcome $k$. The density matrix of the composite system after the feedback control is given by 
\beq
\rho_{\mathrm{f}}^{ABM}(k):=(U^{A}_{k}\otimes U_{k}^{B}) \rho^{ABM}(k) ( U_{k}^{A}\otimes U_{k}^{B})^{\dagger}.
\eeq
The final Hamiltonian of the system (at time $t_{\mathrm{f}}$) depends on $k$ and is given by $H^{A}_{\mathrm{f}}(k)\otimes H^{B}_{\mathrm{f}}(k)$.

4. We erase the information registered in the memory to complete the thermodynamic cycle. We attach a heat bath $R$ to the memory, and perform a unitary operation $U^{MR}_{\text{ers}}$ so that the state of the memory goes back to the standard state, i.e., the support of $\rho^{M}_{\text{ers}}:=\Tr_{R}[\rho^{MR}_{\text{ers}}]$ is in $\bf{H}^{M}_{0}$, where 
\beq
\rho_{\text{ers}}^{MR}=U^{MR}_{\text{ers}}(\rho^{M}_{\mathrm{f}}\otimes\rho^{R})U^{\dagger MR}_{\text{ers}}.
\eeq
Here $\rho^{R}=\rho^{R}_{\text{can}}:=\ee^{-\beta(H^{R}-F^{R})}$ is the initial density operator of the heat bath, and $\rho^{M}_{\mathrm{f}}:=\sum_{k}p_{k}\rho_{\mathrm{f}}^{M}(k)=\sum_{k}p_{k}\Tr_{AB}[\rho_{\mathrm{f}}^{ABM}(k)]$ is the final density operator of the memory which is obtained by taking the ensemble average over $k$. The Hamiltonian and free energy of the heat bath are denoted by $H^{R}$ and $F^{R}$, respectively.

The primary purpose of this paper is to show that the upper bound for the net work gain depends on the amount of correlation that can be utilized via measurement and feedback, as shown in inequality~(\ref{4}). In particular, we can extract work for entangled states beyond classical correlation. For example, the maximum work gain for a 2-qubit system is given by $2\bz T\ln 2$ for entangled states, whereas it is given by $\bz T\ln2$ for classically correlated states~\cite{comment}. This difference arises from the measurement cost as discussed later (see (\ref{2})). To derive~(\ref{4}), we show three inequalities concerning the extractable work  from feedback control~(\ref{1}), the measurement cost~(\ref{2}), and the information-erasure cost~(\ref{3}).

\section{\label{sec:slaw}Generalized second laws}

\subsection{Second law of thermodynamics without feedback control}

We derive second-law-like inequalities using entropy production $\sigma$ which measures the irreversibility of the thermodynamic process. Let us consider a thermodynamic process where the initial density matrix is given by $\rho_{\mathrm{i}}$ and the system undergoes a unitary evolution $U:=\mathrm{T}\exp[-\ii\int^{t_{\mathrm{f}}}_{t_{\mathrm{i}}}H(t)\mathrm{d}t]$, where $H(t)$ is the Hamiltonian of the system. Then, the entropy production is given by
\beq
\sigma:=-\Tr[\rho_{\mathrm{f}}\ln\rho_{\mathrm{r}}]-S(\rho_{\mathrm{i}}), \label{entunitary}
\eeq
where $\rho_{\mathrm{f}}:=U\rho_{\mathrm{i}}U^{\dagger}$ is the final density matrix and $\rho_{\mathrm{r}}$ is a reference state which is the initial state of the backward process (time-reversed protocol) in the context of detailed fluctuation theorems~\cite{Sagawa5}. The entropy production of the system is non-decreasing: $\sigma\geq 0$, which can be evaluated as follows:
\beqa
\sigma&=&-\Tr[\rho_{\mathrm{f}}\ln\rho_{\mathrm{r}}]-S(\rho_{\mathrm{f}}) \nonumber \\
&=& S(\rho_{\mathrm{f}}||\rho_{\mathrm{r}})\geq 0, \label{sigmaslaw}
\eeqa
where the last inequality results from the positivity of the relative entropy: $S(\rho||\rho'):= -\Tr[\rho\ln\rho']-S(\rho)\geq 0$~\cite{Nielsen}. Although the inequality~(\ref{sigmaslaw}) holds for any choice of reference states, we assume that the reference state is given by the canonical distribution $\rho_{\mathrm{r}}=\ee^{-\beta(H(t_{\mathrm{f}})-F_{\mathrm{f}})}$ since the entropy production can be related to thermodynamic quantities. In fact, if the initial state is given by the canonical distribution, i.e., $\rho_{\mathrm{i}}=\ee^{-\beta(H(t_{\mathrm{i}})-F_{\mathrm{i}})}$, the entropy production is given by  
\beq
\sigma=-\beta(W_{\text{ext}}+\Delta F),
\eeq
where $W_{\text{ext}}:=\Tr[\rho_{\mathrm{i}}H(t_{\mathrm{i}})]-\Tr[\rho_{\mathrm{f}}H(t_{\mathrm{f}})]$ is the extractable work, $\Delta F:=F_{\mathrm{f}}-F_{\mathrm{i}}$ is the free-energy difference, and $\beta$ is the inverse temperature. Thus the non-negativity of the entropy production leads to the conventional second law: 
\beq
W_{\text{ext}}\leq -\Delta F.
\eeq

\subsection{Extractable work via feedback control}

When information processing is involved, the second law should be generalized by including information contents~\cite{Sagawa1,Sagawa2,Sagawa3}. The relevant information content is the information gain~\cite{Ozawa,Groenewold,Sagawa1}, which characterizes the additional knowledge about the system acquired by the measurement:
\beq
I(\rho_{\mathrm{i}}^{A}:X^{M}):= S(\rho_{\mathrm{i}}^{A})-\sum_{k}p_{k}S(\rho^{A}(k)). \label{informationgain}
\eeq
The information gain is bounded from above by $I(\rho_{\mathrm{i}}^{A}:X^{M})\leq H(X^{M})$, where $H(X^{M}):=-\sum_{k}p_{k}\ln p_{k}$ is the Shannon entropy. Equation~(\ref{informationgain}) may take negative values in general due to the measurement back action. However, it can be shown that the information gain is non-negative:
\beq
0\leq I(\rho_{\mathrm{i}}^{A}:X^{M})\leq H(X^{M}), \label{infoqc}
\eeq
if the postmeasurement state~(\ref{postmesstateab}) is expressed by a single measurement operator $M^{A}_{k}$, i.e., $\rho^{AB}(k)=p_{k}^{-1}M^{A}_{k}\rho^{AB}_{\mathrm{i}}M^{\dagger A}_{k}$~\cite{Ozawa,Sagawa1}. In this paper, we consider the situation in which the information gain satisfies the inequality~(\ref{infoqc}) so that we can utilize the measurement result.

We define the following quantity 
\beq
\sigma^{A}_{\text{ext}}:=-\sum_{k}p_{k}\Tr[\rho_{\mathrm{f}}^{A}(k)\ln\rho^{A}_{\mathrm{r}}(k)]-S(\rho^{A}_{\mathrm{i}}), \label{entdefa}
\eeq
where $\rho^{A}_{\mathrm{r}}(k)$ is the reference state of $A$ depending on $k$. It is similar to the entropy production defined in Eq.~(\ref{entunitary}), but $\sigma^{A}_{\text{ext}}$ can take negative values since it contains not only the dissipative part but also the information contents via measurement and feedback:
\beqa
\sigma^{A}_{\text{ext}}
&=&\sum_{k}p_{k}S(\rho_{\mathrm{f}}^{A}(k)||\rho_{\mathrm{r}}^{A}(k))-S(\rho^{A}_{\mathrm{i}})+\sum_{k}p_{k}S(\rho_{\mathrm{f}}^{A}(k)) \nonumber \\
& \geq& -I(\rho_{\mathrm{i}}^{A}:X^{M}), \label{ent1a}
\eeqa
where the last inequality follows from the positivity of the relative entropy and $S(\rho^{A}_{\mathrm{f}}(k))=S(\rho^{A}(k))$. The equality in (\ref{ent1a}) holds when $\rho_{\mathrm{f}}^{A}(k)=\rho_{\mathrm{r}}^{A}(k)$. From inequality (\ref{ent1a}), we observe that  $\sigma^{A}_{\text{ext}}+I(\rho_{\mathrm{i}}^{A}:X^{M})$ is nonnegative and measures the irreversibility of the thermodynamic process. It also shows that the entropy of $A$ can be decreased up to the information gain $I(\rho_{\mathrm{i}}^{A}:X^{M})$. Thus $\sigma^{A}_{\text{ext}}$ measures the amount of entropy that can be decreased from the system via measurement and feedback.

 A similar relation holds for the subsystem $B$:
\beq
\sigma^{B}_{\text{ext}}\geq  -I(\rho_{\mathrm{i}}^{B}:X^{M}), \label{ent1b}
\eeq
where 
\beq
\sigma^{B}_{\text{ext}}:=-\sum_{k}p_{k}\Tr[\rho_{\mathrm{f}}^{B}(k)\ln\rho^{B}_{\mathrm{r}}(k)]-S(\rho^{B}_{\mathrm{i}})
\eeq
is the amount of entropy that can be decreased via measurement and feedback on $B$, $\rho^{B}_{\mathrm{r}}(k)$ is the reference state of $B$ depending on $k$, and 
\beq
I(\rho_{\mathrm{i}}^{B}:X^{M}):=S(\rho^{B}_{\mathrm{i}})-\sum_{k}p_{k}S(\rho^{B}(k)) \label{chi}
\eeq is the Holevo $\chi$ quantity which gives the amount of information about $B$ acquired by the measurement. Apart from the information gain~(\ref{informationgain}), Eq.~(\ref{chi}) is nonnegative for any measurement operator $M^{A}_{k,a,b}$~\cite{Nielsen}:
\beq
0\leq I(\rho_{\mathrm{i}}^{B}:X^{M})\leq H(X^{M}), 
\eeq
since the effect of measurement back action does not directly affect on $B$. 

Next, we derive the upper bound of the work that can be extracted from subsystems $A$ and $B$ by assuming that their initial states are given by the canonical distributions $\rho^{A}_{\mathrm{i}}=\ee^{\beta(F^{A}_{\mathrm{i}}-H^{A}_{\mathrm{i}})}$ and $\rho^{B}_{\mathrm{i}}=\ee^{\beta(F^{B}_{\mathrm{i}}-H^{B}_{\mathrm{i}})}$. Such conditions are met by the following entangled state~\cite{squeezed}: 
\beq
\ket{\psi^{AB}}=Z^{-\half}\sum_{k}\ee^{-\beta \epsilon_{k}/2}\ket{k}_{A}\otimes\ket{k}_{B},
\eeq
where $\ket{k}_{A,B}$ is the $k$-th energy eigenstate and $\epsilon_{k}$ is the energy eigenvalue of $A$ and $B$, respectively, and $Z^{-\half}$ is the normalization constant. Such a quantum correlated state can be created experimentally by parametric amplification~\cite{amplification}, and it plays a pivotal role in such diverse phenomena as Hawking radiation~\cite{Hawking}, the dynamical Casimir effect~\cite{Moore,Fulling}, and the Unruh effect~\cite{Unruh}. By choosing the reference state to be the canonical distribution $\rho^{A}_{\mathrm{r}}(k)=\ee^{\beta(F^{A}_{\mathrm{f}}(k)-H^{A}_{\mathrm{f}}(k))}$, where $F^{A}_{\mathrm{f}}(k):=-\beta^{-1}\ln\Tr\exp[-\beta H^{A}_{\mathrm{f}}(k)]$, we can relate $\sigma^{A}_{\text{ext}}$ to work done on the system $A$
\beq
W^{A}_{\text{ext}}:= \Tr[ \rho_{\mathrm{i}}^{A}H^{A}_{\mathrm{i}}]-\sum_{k}p_{k}\Tr[\rho_{\mathrm{f}}^{A}(k)H^{A}_{\mathrm{f}}(k)]
\eeq
and the free-energy difference $\Delta F^{A}:=\sum_{k}p_{k}F_{\mathrm{f}}^{A}(k)-F_{\mathrm{i}}^{A}$ as 
\beq
\sigma^{A}_{\text{ext}} = -\beta W^{A}_{\text{ext}}-\beta \Delta F^{A}.
\eeq
A similar relation holds for subsystem $B$. Then, (\ref{ent1a}) and (\ref{ent1b}) lead to the inequality concerning the extractable work from the quantum correlated state:
\beqa
W_{\text{ext}}^{A}+W_{\text{ext}}^{B} &\leq& -\Delta F^{A}-\Delta F^{B} \nonumber \\ & +& \bz T\left[I(\rho_{\mathrm{i}}^{A}:X^{M})+I(\rho_{\mathrm{i}}^{B}:X^{M})\right]. \label{1}
\eeqa
The work gain beyond the conventional second law is expressed by the mutual information contents between each subsystem and the memory: $I(\rho^{A}:X^{M})$ and $I(\rho^{B}:X^{M})$. For initially uncorrelated states, $I(\rho_{\mathrm{i}}^{B}:X^{M})=0$ because the work gain of $B$ arises from the initial correlation between the subsystems. Then, the result of Ref.~\cite{Sagawa1} is reproduced for $A$: 
\beq
W_{\text{ext}}^{A}\leq -\Delta F^{A}+\bz TI(\rho_{\mathrm{i}}^{A}:X^{M}),
\eeq
and the conventional second law holds for $B$: $W_{\text{ext}}^{B}\leq -\Delta F^{B}$. For a single system, a feedback control that achieves the upper bound of the extractable work is constructed in Ref.~\cite{Jacobs}.

\subsection{Measurement cost}

Next, we derive inequalities for the measurement process by following Ref.~\cite{Sagawa2}, which gives the lower bound of the measurement cost. We define the following quantity that measures the additional entropy produced via measurement: 
\beq
\sigma^{M}_{\text{mes}}=-\Tr[\rho^{M}_{\mathrm{f}}\ln\rho^{M}_{\mathrm{r}}]-S(\rho^{M}_{0}),
\eeq
where $\rho^{M}_{\mathrm{r}}=\sum_{k}p_{k}\rho^{M}_{\mathrm{r}}(k)$ is the reference state of the memory and the support of $\rho^{M}_{\mathrm{r}}(k)$ belongs to $\bf{H}^{M}_{k}$. Since the supports of $\rho^{M}_{k}$ and $\rho^{M}_{\mathrm{r}}(k)$ belong to the mutually orthogonal subspace $\bf{H}^{M}_{k}$, we have 
\beq
-\Tr[\rho^{M}_{\mathrm{f}}\ln\rho^{M}_{\mathrm{r}}]=-\sum_{k}p_{k}\Tr[\rho^{M}_{\mathrm{f}}(k)\ln\rho^{M}_{\mathrm{r}}(k)]+H(X^{M}). \label{memoryassum}
\eeq
Now we derive the lower bound of $\sigma^{M}_{\text{mes}}$. First, note that
\beqa
& &\sigma^{M}_{\text{mes}}-I(\rho_{\mathrm{i}}^{AB}:X^{M}) \nonumber \\
&=& S(\rho^{ABM}_{\mathrm{f}})-S(\rho^{ABM}_{\mathrm{i}})-\sum_{k}p_{k}S(\rho^{ABM}_{\mathrm{f}}(k)) \nonumber \\ 
&+&\sum_{k}p_{k}S(\rho^{AB}_{\mathrm{f}}(k))        -\sum_{k}p_{k}\Tr[\rho^{M}_{\mathrm{f}}(k)\ln\rho^{M}_{\mathrm{r}}(k)], \label{assumpa}
\eeqa
where $\rho^{ABM}_{\mathrm{f}}:=\sum_{k}p_{k}\rho^{ABM}_{\mathrm{f}}(k)$. The equality in Eq.~(\ref{assumpa}) follows from Eq.~(\ref{memoryassum}) and the fact that $\rho^{ABM}_{\mathrm{f}}(k)$'s are mutually orthogonal. We then have
\beqa
& &\sigma^{M}_{\text{mes}}-I(\rho_{\mathrm{i}}^{AB}:X^{M}) \nonumber \\
&=& \Delta S + \sum_{k}p_{k}S(\rho^{ABM}_{\mathrm{f}}(k)|| \rho^{AB}_{\mathrm{f}}(k)\otimes\rho^{M}_{\mathrm{r}}(k))\nonumber \\
&\geq& 0     \label{mes1},
\eeqa
where $\Delta S:=S(\rho^{ABM}_{\mathrm{f}})-S(\rho^{ABM}_{\mathrm{i}})$ and the last inequality is satisfied because the relative entropy is positive and the von Neumann entropy of the density matrix that is averaged over many runs of the projective measurement does not decrease in comparison with the entropy of the premeasurement state~\cite{Nielsen}. The equality in (\ref{mes1}) is satisfied if $\rho^{ABM}_{\mathrm{f}}(k)=\rho^{AB}_{\mathrm{f}}(k)\otimes\rho^{M}_{\mathrm{r}}(k)$ and $\Delta S =0$. Inequality (\ref{mes1}) gives the fundamental lower bound of entropy that is produced under measurement. The bound is expressed by the information gain between the total system and the memory: 
\beq
I(\rho_{\mathrm{i}}^{AB}:X^{M}):=S(\rho_{\mathrm{i}}^{AB})-\sum_{k}p_{k}S(\rho^{AB}(k)).
\eeq
When the initial and reference density operators of the memory obey the canonical distributions $\rho^{M}_{0}=\rho^{M}_{0,\text{can}}:=\ee^{\beta(F^{M}_{0}-H^{M}_{0})}$ and $\rho^{M}_{\mathrm{r}}(k)=\rho^{M}_{k,\text{can}}:=\ee^{\beta(F^{M}_{k}-H^{M}_{k})}$, $\sigma^{M}_{\text{mes}}$ can be related to work as 
\beq
\sigma^{M}_{\text{mes}}= H(X^{M})+\beta W^{M}_{\text{mes}}-\beta\Delta F^{M},
\eeq
where $W^{M}_{\text{mes}}:=\sum_{k}p_{k}\Tr[\rho_{k}^{M}H_{k}^{M}]-\Tr[\rho^{M}_{0}H_{0}^{M}]$ and $\Delta F^{M}:=\sum_{k}p_{k}F^{M}_{k}-F^{M}_{0}$. Then, (\ref{mes1}) reduces to
\beq
W_{\text{mes}}^{M}\geq \bz T[I(\rho_{\mathrm{i}}^{AB}:X^{M})-H(X^{M})]+\Delta F^{M}, \label{2}
\eeq
which gives the lower bound of the measurement cost. For initially uncorrelated states, $I(\rho_{\mathrm{i}}^{AB}:X^{M})$ reduces to the classical result $I(\rho_{\mathrm{i}}^{A}:X^{M})$~\cite{Sagawa2}. The measurement cost depends on the amount of accessible information for the feedback controller, and in the present case, the nonlocal correlation between the subsystems should be considered.

\subsection{Erasure cost}
Next, we derive the lower bound of the work cost for initializing the memory. The entropy production is given by
\beq
\sigma^{M}_{\text{ers}}=S(\rho^{M}_{\text{ers}})-S(\rho^{M}_{\mathrm{f}})-\beta Q_{\text{ers}},
\eeq
where $Q_{\text{ers}}:=\Tr[H^{R}\rho^{R}_{\mathrm{f}}]-\Tr[H^{R}\rho^{R}_{\text{ers}}]$. Since the erasure process is unitary for $MR$, the entropy production is nonnegative: 
\beqa
\sigma^{M}_{\text{ers}}&=&S(\rho^{M}_{\mathrm{f}})-S(\rho^{M}_{\mathrm{f}})-\Tr[\rho^{R}_{\text{ers}}\ln\rho^{R}_{\text{can}}]-S(\rho^{R}) \nonumber \\
&=&-S(\rho^{R}_{\text{ers}})+S(\rho^{M}_{\text{ers}})-\Tr[\rho^{R}_{\text{ers}}\ln\rho^{R}_{\text{can}}]\nonumber \\
&=&-\Tr[\rho^{MR}_{\text{ers}}\ln\rho^{M}_{\text{ers}}\otimes\rho^{R}_{\text{can}}]-S(\rho^{MR}_{\text{ers}}) \nonumber \\
&=&S(\rho^{MR}_{\text{ers}}||\rho^{M}_{\text{ers}}\otimes\rho^{R}_{\text{can}})\geq 0.
\eeqa
If $\rho^{M}_{\mathrm{f}}=\sum_{k}p_{k}\rho^{M}_{k,\text{can}}$ and $\rho^{M}_{\text{ers}}=\rho^{M}_{0,\text{can}}$, we have 
\beq
\sigma^{M}_{\text{ers}}=-H(X^{M})+\beta W^{M}_{\text{ers}}+\beta \Delta F^{M},
\eeq
where 
\beq
W^{M}_{\text{ers}}:=\Tr[\rho_{\text{ers}}^{M}H_{0}^{M}]-\sum_{k}p_{k}\Tr[\rho^{M}_{k,\text{can}}H^{M}_{k}]-Q_{\text{ers}}.
\eeq
Then, the erasure cost is given by~\cite{Sagawa2}:
\beq
W_{\text{ers}}^{M}\geq \bz TH(X^{M})-\Delta F^{M}. \label{3}
\eeq
For $\Delta F^{M}=0$, this inequality reduces to Landauer's principle~\cite{Landauer}.

\begin{figure}[h]
\includegraphics[width=.4\textwidth]{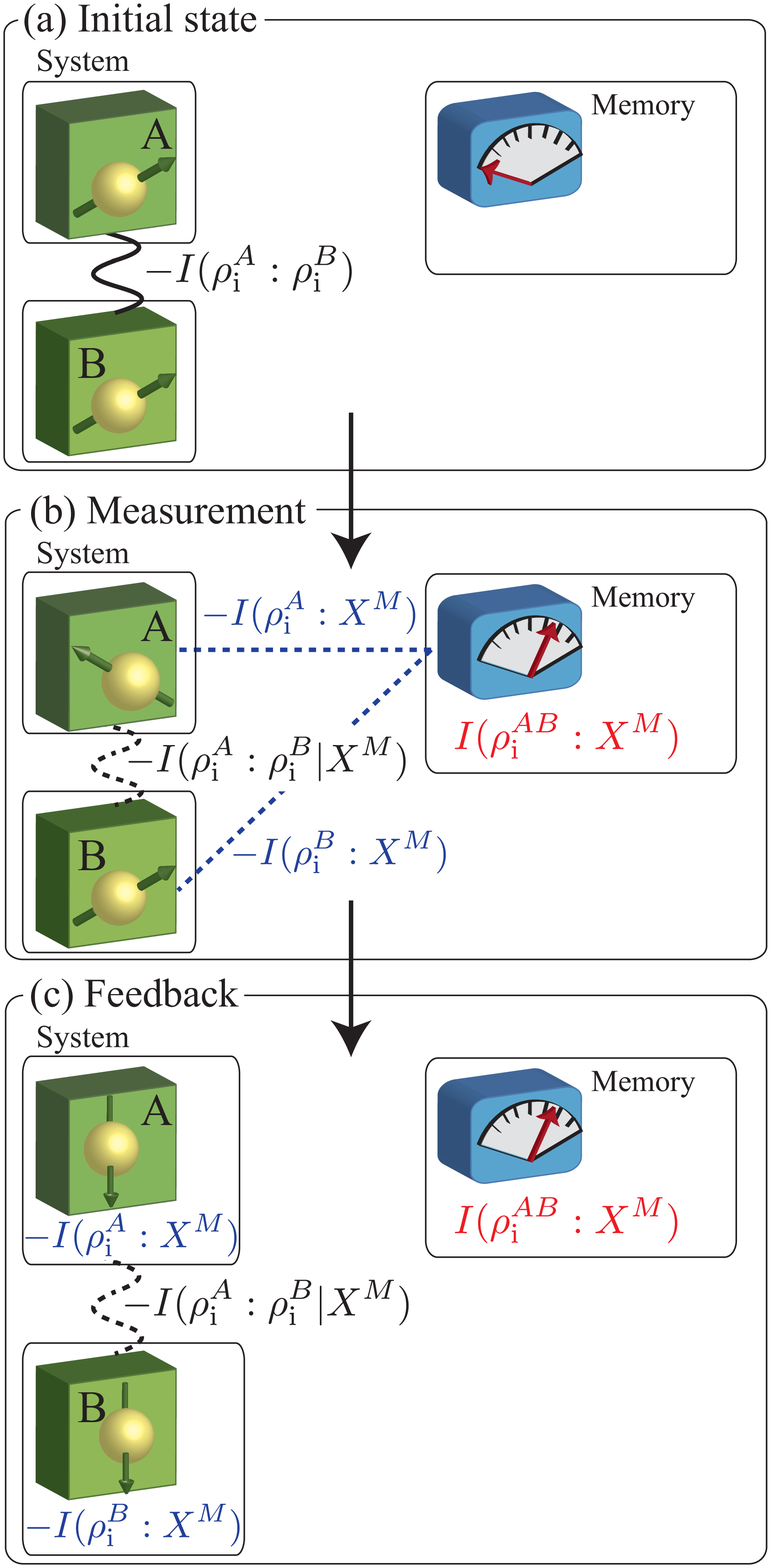}
\caption{\label{fig:wide}(color online). Entropy transfer in the protocol: (a) $A$ and $B$ are initially entangled, where the negative entropy is described by the quantum mutual information $-I(\rho_{\mathrm{i}}^{A}:\rho_{\mathrm{i}}^{B})$. We can decrease the entropy of $A$ and $B$ up to $I(\rho_{\mathrm{i}}^{A}:\rho_{\mathrm{i}}^{B})$ by performing an appropriate unitary transformation on the entire system. (b) Measurement on $A$ and entropy transfer from the system to $M$. The information gain of subsystem $A$ is given by $-I(\rho_{\mathrm{i}}^{A}:X^{M})$ and that of subsystem $B$ by $-I(\rho_{\mathrm{i}}^{B}:X^{M})$, whereas the memory generates a positive entropy $I(\rho_{\mathrm{i}}^{AB}:X^{M})$. (c) Feedback control performed on both $A$ and $B$ produce negative entropies $-I(\rho_{\mathrm{i}}^{A}:X^{M})$ and $-I(\rho_{\mathrm{i}}^{B}:X^{M})$, which arise from the initial correlation $-I(\rho_{\mathrm{i}}^{A}:\rho_{\mathrm{i}}^{B})$ and information gain $-I(\rho_{\mathrm{i}}^{AB}:X^{M})$. Consequently, the total entropy of $A$ and $B$ decreases, and work can be extracted from them. The correlation $I(\rho_{\mathrm{i}}^{A}:\rho_{\mathrm{i}}^{B}|X^{M})$ between $A$ and $B$ remains nonvanishing, which, however, cannot be utilized unless $A$ and $B$ are brought together to perform an appropriate unitary transformation on the entire state.}
\end{figure}

\subsection{Net work gain from entangled states}
By combining the results derived above, we give the lower bound of the net work gain from initially entangled states. The net entropy that can be decreased from the total system is characterized by $\sigma_{\text{net}}=\sigma_{\text{ext}}^{A}+\sigma_{\text{ext}}^{B}+\sigma_{\text{mes}}^{M}$, and it satisfies the following inequality:
\beq
\sigma_{\text{net}}\geq -[I(\rho_{\mathrm{i}}^{A}:\rho_{\mathrm{i}}^{B})-I(\rho_{\mathrm{i}}^{A}:\rho_{\mathrm{i}}^{B}|X^{M})] , \label{entnet}
\eeq
where $I(\rho^{A}_{\mathrm{i}}:\rho^{B}_{\mathrm{i}})$ is the quantum mutual information given by Eq.~(\ref{quantummutual}) and 
\beqa
& &I(\rho_{\mathrm{i}}^{A}:\rho_{\mathrm{i}}^{B}|X^{M}) \nonumber \\:=& &\sum_{k}p_{k} [S(\rho^{A}(k))+S(\rho^{B}(k))-S(\rho^{AB}(k))]
\eeqa
is the quantum mutual information between the subsystems conditioned upon the measurement outcomes $k$'s. Inequality~(\ref{entnet}) can be derived by combining (\ref{ent1a}), (\ref{ent1b}), and (\ref{mes1}), and using the identity: 
\beqa
& &I( \rho_{\mathrm{i}}^{A}: X^{M}) +I(\rho_{\mathrm{i}}^{B}:X^{M})-I(\rho_{\mathrm{i}}^{AB}:X^{M}) \nonumber \\ 
&=& I(\rho_{\mathrm{i}}^{A}:\rho_{\mathrm{i}}^{B})-I(\rho_{\mathrm{i}}^{A}:\rho_{\mathrm{i}}^{B}|X^{M}),
\eeqa
which results from the entropy balance in the protocol (see Fig.~\ref{fig:wide}). We can decrease the entropy of the total system up to the right-hand side of (\ref{entnet}) with the help of the measurement and feedback.

We define the net work gain $W_{\text{net}}=W^{A}_{\text{ext}}+W^{B}_{\text{ext}}-W^{M}_{\text{mes}}-W^{M}_{\text{ers}}$, which describes the work gain due solely to the initial correlation. It follows from (\ref{1}), (\ref{2}), and (\ref{3}) that
\beq
W_{\text{net}} \leq \bz T\left[I(\rho_{\mathrm{i}}^{A}:\rho_{\mathrm{i}}^{B})-I(\rho_{\mathrm{i}}^{A}:\rho_{\mathrm{i}}^{B}|X^{M})\right] -\Delta F^{AB}. \label{4}
\eeq
This is the primary result of this paper. The net work gain depends on the amount of the initial correlation between the subsystems that the memory can access. This can be seen if we use the Venn diagram, since the term $I(\rho_{\mathrm{i}}^{A}:\rho_{\mathrm{i}}^{B})-I(\rho_{\mathrm{i}}^{A}:\rho_{\mathrm{i}}^{B}|X^{M})$ expresses the information that is shared by the three states (i.e., states of $A$, $B$ and $M$). Since the available correlation for the quantum state is larger than that for the classical state, we can extract work from the entangled state beyond classical correlation. For a two-qubit system, the maximum value of the quantum mutual information $I(\rho^{A}_{\mathrm{i}}:\rho^{B}_{\mathrm{i}})$ is $2\ln2$ for entangled states whereas it is $\ln 2$ for classically correlated states. If we choose an appropriate scheme for measurement and feedback control which utilize all the correlation between subsystems, i.e., $I(\rho^{A}_{\mathrm{i}}:\rho^{B}_{\mathrm{i}}|X^{M})=0$, we observe that the upper bound of the net work gain for the entangled states is twice as large as that of the classically correlated states. The result~(\ref{4}) demonstrates that quantum correlation can be utilized as a resource to obtain the thermodynamic work gain. For initially uncorrelated states, inequality~(\ref{4}) reduces to the conventional second law for the total system including the memory: $W_{\text{net}}\leq -\Delta F$. By considering the cost for establishing the correlation, we can show that~(\ref{4}) is consistent with the conventional second law as shown below.

\subsection{Cost for creating entanglement}

Let us start with an uncorrelated state $\rho^{A}\otimes\rho^{B}$ of the composite system $AB$, and perform a unitary transformation $U^{AB}$ to establish correlation: $\rho^{AB}_{\mathrm{i}}=U^{AB}\rho^{A}\otimes\rho^{B}U^{\dagger AB}$. The amount of entropy that is produced in subsystems $A$ and $B$ during this process is given by
\beqa
\sigma^{A}_{\text{ent}}&:=& -\Tr[\rho^{A}_{\mathrm{i}}\ln \rho^{A}_{\mathrm{r}}] -S(\rho^{A}), \nonumber \\
\sigma^{B}_{\text{ent}}&:=&-\Tr[ \rho^{B}_{\mathrm{i}}\ln \rho^{B}_{\mathrm{r}}] -S(\rho^{B}) ,
\eeqa
where the reference states are given by $\rho^{A}_{\mathrm{r}}$ and $\rho^{B}_{\mathrm{r}}$. By noting that $S(\rho^{A})+S(\rho^{B})=S(\rho^{AB}_{\mathrm{i}})$, we obtain
\beqa
\sigma^{A}_{\text{ent}}+\sigma^{B}_{\text{ent}} &=& S(\rho^{A}_{\mathrm{i}}||\rho^{A}_{\mathrm{r}}) +S(\rho^{B}_{\mathrm{i}}||\rho^{B}_{\mathrm{r}}) \nonumber \\ & &+S(\rho^{A}_{\mathrm{i}})+S(\rho^{B}_{\mathrm{i}})-S(\rho^{A})-S(\rho^{B}) \nonumber \\
&\geq& I(\rho^{A}_{\mathrm{i}}:\rho^{B}_{\mathrm{i}}) \label{entcost}.
\eeqa
The inequality~(\ref{entcost}) states that the lower bound of creating correlation is given by the quantum mutual information between subsystems. The equality in~(\ref{entcost}) is achieved if and only if $\rho^{A}_{\mathrm{r}}=\rho^{A}_{\mathrm{i}}$ and $\rho^{B}_{\mathrm{r}}=\rho^{B}_{\mathrm{i}}$. Combining inequalities~(\ref{entnet}) and (\ref{entcost}), we obtain the lower bound on the total entropy that is produced under a thermodynamic cycle as
\beq
\sigma_{\text{cycle}}\geq I(\rho^{A}_{\mathrm{i}}:\rho^{B}_{\mathrm{i}}|X^{M})\geq 0, \label{cycle}
\eeq
where $\sigma_{\text{cycle}}:=\sigma^{A}_{\text{ent}}+\sigma^{B}_{\text{ent}}+\sigma_{\text{net}}$. From~(\ref{cycle}), we find that the entropy production is always positive, and bounded from below by $I(\rho^{A}_{\mathrm{i}}:\rho^{B}_{\mathrm{i}}|X^{M})$. The additional entropy production $I(\rho^{A}_{\mathrm{i}}:\rho^{B}_{\mathrm{i}}|X^{M})$ arises since the feedback control is limited to local unitary operations, i.e. $U^{A}_{k}\otimes U^{B}_{k}$. If we are allowed to perform a unitary operation $U^{AB}_{k}$ on the composite system for the feedback control, we can show that the entropy produced under a thermodynamic cycle restores the conventional second law. First, note that the entropy produced on the composite system is bounded by
\beq
\sigma'^{AB}_{\text{ext}}\geq -I(\rho^{AB}_{\mathrm{i}}:X^{M}), \label{compositegain}
\eeq
where 
\beq
\sigma'^{AB}_{\text{ext}}:=-\sum_{k}p_{k}\Tr[\rho^{AB}(k)\ln\rho^{AB}_{\mathrm{r}}(k)]-S(\rho^{A}\otimes\rho^{B}),
\eeq
and $\rho^{AB}_{\mathrm{r}}(k)$ is the reference state which depends on $k$. We note that this inequality directly follows by applying the result discussed in Ref.~\cite{Sagawa1} to the composite system $AB$. Combined with the measurement cost~(\ref{mes1}), we observe that the total entropy produced for a cyclic process restores the conventional second law:
\beq
\sigma'_{\text{cycle}}\geq 0, \label{compositecycle}
\eeq
where $\sigma'_{\text{cycle}}:=\sigma'^{AB}_{\text{ext}}+\sigma^{M}_{\text{mes}}$. Comparing inequalities~(\ref{cycle}) and (\ref{compositecycle}), we find that $I(\rho^{A}_{\mathrm{i}}:\rho^{B}_{\mathrm{i}}|X^{M})$ measures the difference in the performance between local and nonlocal feedback operations.

To address the question of whether a more general LOCC protocol would leads to a better work gain, let us prepare another memory that measures B depending on A's measurement outcome $k$. Suppose that the outcome is $l\in Y^{M}$, and the post-measurement state is given by 
\beq
\rho^{AB}(k,l)=\frac{1}{p_{l|k}}M^{B}_{k,l}\rho_{\mathrm{f}}^{AB}(k)M^{B\dagger}_{k,l},
\eeq
where $M^{B}_{k,l}$ is the measurement operator which depends on $k$ and satisfy the relation $\sum_{l}M^{\dagger B}_{k,l}M^{B}_{k,l}=1$. The measurement on B enables us to obtain further knowledge about A and B: 
\beq
I(\rho_{\mathrm{f}}^{A(B)}(k):Y^{M}):=S(\rho_{\mathrm{f}}^{A(B)}(k))-\sum_{l}p_{l|k}S(\rho^{A(B)}(k,l)).
\eeq
Thus the results in (\ref{1})-(\ref{4}) can be generalized by replacing $I(\rho_{\mathrm{i}}^{A}:X^{M})$ with 
\beqa
I(\rho_{\mathrm{i}}^{A}:X^{M}Y^{M})&:=&S(\rho_{\mathrm{i}}^{A})-\sum_{k,l}p_{l|k}p_{k}S(\rho^{A}(k,l)) \nonumber \\
&=&I(\rho_{\mathrm{f}}^{A}(k):Y^{M})+I(\rho^{A}_{\mathrm{i}}:X^{M}),
\eeqa
 and similar generalization can be made for $B$ and $AB$. Thus, (\ref{4}) is generalized to 
\beqa
W_{\text{net}} &\leq& \bz T[I(\rho^{A}_{\mathrm{i}}:\rho^{B}_{\mathrm{i}})-\sum_{k,l}p_{l|k}p_{k}I(\rho^{A}(k,l):\rho^{B}(k,l))] \nonumber \\
& &-\Delta F^{AB},
\eeqa
which results in a better work gain. The generalization to a more complicated LOCC protocol is straightforward.

\section{\label{sec:discussion}Discussions and conclusion}

In contrast to Refs.~\cite{Zurek,Brodutch}, we have added here the work gain which results solely from entanglement. We also note that the work gain from entangled states using LOCC protocols was discussed in Refs.~\cite{Horodecki,Oppenheim}. Our result is quantitatively different from theirs because the upper bound of work gain in these references gives the same value for both entangled states and classically correlated states for the same marginal density matrix, whereas our result shows that genuine quantum entanglement can be utilized to achieve further work gain. Further work gain arises from the difference in the measurement cost because the initial entropy of $AB$ is smaller for quantum correlated states than for classical ones. Therefore, the cost required for the memory to compensate for the decrease in the composite system's entropy via feedback is different. It is an interesting future problem to apply quantum discords~\cite{discord1,discord2} to our setup by comparing thermodynamic work gain between global operations and LOCC protocols. It is also an interesting future problem to construct an explicit model which achieves the lower bound of the measurement cost~(\ref{2}) and also the upper bound of the net work gain~(\ref{4}) from entangled states.

In conclusion, we have shown that entanglement can be utilized to extract work beyond classical correlation, and that the maximal work is determined by the difference in the quantum mutual information between the subsystems that expresses the memory's accessible information about the system. The results of our work serve as the foundations for controlling quantum-correlated thermodynamic systems and set the fundamental upper bound on the amount of work gain that can be obtained from such systems.

This work was supported by Grants-in-Aid for Scientific Research (Kakenhi Nos. 22340114, 22103005, 24840015, 25287098, and 254105), Global COE Program ``The Physical Sciences Frontier," and the Photon Frontier Netowork Program of MEXT of Japan.


\begin{thebibliography}{99}
\bibitem{comment} This bound is asymptotically achieved in the high-temperature limit.
\end{thebibliography}
\end{document}